# Extending high value components performances with Additive Manufacturing: application to naval applications

RAUCH Matthieu[1, 3, a *], PECHET Gatien[1, 3, b], HASCOET Jean-Yves[1, 3, c] and RUCKERT Guillaume[2, 3, d]

[1]Centrale Nantes / GeM - UMR CNRS 6183, 1 rue de la Noe, 44321 Nantes

[2]Naval Group Research, Technocampus Ocean, 5 Rue de L'Halbrane, 44340 Bouguenais

[3]Joint Laboratory of Marine Technology (JLMT) Centrale Nantes - Naval Group

[a]matthieu.rauch@ec-nantes.fr, [b]gatien.pechet@ec-nantes.fr, [c]jean-yves.hascoet@ec-nantes.fr, [d]guillaume.ruckert@naval-group.com

**Abstract.** Additive Manufacturing (AM), consists of depositing material in successive layers to obtain the desired part. The parts produced by AM can thus adopt geometries inaccessible by conventional manufacturing means, for example hollow or lattice structures which considerably reduce their weight while keeping or even improving their mechanical properties. Among the many existing processes, Wire Arc Additive Manufacturing (WAAM) is particularly well suited to the manufacture of large metallic parts. It is characterized by a supply of heat in the form of an electric arc (produced by a welding generator) and a supply of material in the form of wire. This paper will discuss the impact of additive manufacturing to enhance the performances of high value components, based on naval application: the manufacturing of a hollow propeller blade demonstrator of 1.5 m high realized in the laboratory.

## Introduction

The rise of Additive Manufacturing (AM) offers new opportunities such as cost reductions and freedom of manufacturing, depending on the type of the component. Among the AM processes referenced by ISO/ASTM 52900 [1], Direct Energy Deposition (DED) processes build the components layer by layer by feeding the material closed to a CNC controlled heat source effector: Laser Metal Deposition is more dedicated to challenging geometries, functional graded materials parts [2] or repair [3] whereas Wire Arc Additive Manufacturing (WAAM) is more suitable for building large structural parts [4]. These processes are becoming viable to propose technological efficient and cost-effective manufacturing solutions for metallic industrial parts. In contrast with usual manufacturing processes, such as casting or forging.

Hence, the WAAM process uses a high performance welding generator a heat source and wire as feedstock, which provides several advantages: high deposition rate, cost competitiveness of the equipment, large working envelop, etc. It has consequently shown its efficiency for various high-dimensional parts in various materials such as steel, aluminium, titanium [5], [6].

This paper will discuss the manufacture of large scale components by WAAM process cannot only improve the cost and delay aspects but also dramatically increase its performances. A container-ship propeller is used as a case study, which is usually cast. The process requirements and associated tooling and equipment usually constrain the outer forms, ask for the realization of massive and heavy parts. In addition, manufacturing cycles and equipment availability leads to long delivery time.

In recent years, few researches has already focused on the manufacturing of naval propellers by WAAM [7], [8]. The manufacturing potential has been highlighted, but the benefits have been limited by constraining to the initial cast geometry.

H2020 European project RAMSSES [9] objectives is to develop, validate and integrate maritime parts and processes for the new generation of ships. The objective is to realize a proof of concept of significant efficiency improvements by propellers made with AM.

The objective of this research is to demonstrate how additive manufacturing process can extend the performances of high value components, illustrated by a propeller blade demonstrator. The first section of this paper will explain the interest of additive manufacturing, followed by the advantages of multiple axial redundancies illustrated by practical implementation and finally the evaluation of the obtained hollow blade.

## Evaluation of AM interest

**High performance propeller geometry enabled.** A major interest of the additive manufacturing is to handle design complexity with lower cost. The components can be reshaped from their functional requirement, overcoming some constraints due to traditional manufacturing process. For the marine propeller, its optimal geometry was designed from fluid mechanics. Then, based on the structural mechanics behavior to define external geometry and minimal thickness, it was possible to leave a cavity inside and consequently propose a hollow blade design [10]. Hollow blades can effectively reduce noise and vibrations of the propellers, thus beneficial to the marine wildlife. Also, it improves their hydrodynamic efficiency by reducing cavitation phenomenon, resulting with interesting economic impacts. The Fig. **1** illustrates the geometry of the container-ship propeller of 7.5 m diameter. For proof of concept by WAAM process, only a hollow blade is manufactured with a downscaling of 0.5 to fit the available manufacturing cell of Ecole Centrale laboratory.

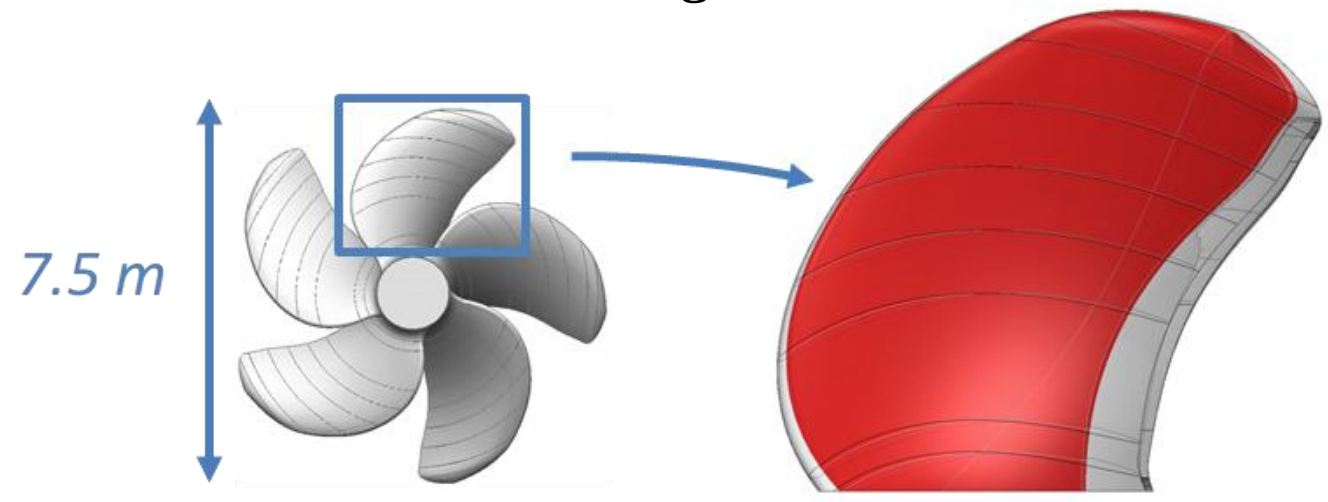

Fig. 1 Illustration of the monobloc propeller with hollow blades.

**DFAM to lever geometric complexities.** Despite the capability of AM to manufacture complex parts, there are few process constraints. Hence some geometric features can be modified to improve the process technical and economic efficiency. As the material is provided locally, toolpaths parameters can be adapted locally to reach the objective. This process is called DFAM (Design For Additive Manufacturing). For instance, the leading edge of a propeller blade has usually a tip shape. This shape is not well adapted to the WAAM process, because the significant heat supply tends to cause a collapse when the molten pool comes at this area. To reduce this effect, an extra thickness is added locally (Fig. **2**). This geometry change aims to add thermal inertia at this location and to avoid the deposited material from collapsing. This extra thickness will later be eliminated by post-production machining.

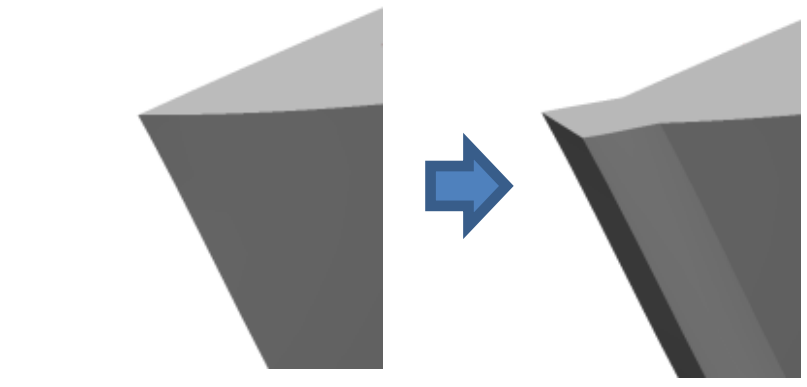

Fig. 2 DFAM of the trailing edge of the blade

**Deposited material properties.** Additive manufacturing provides a better control of the local characteristics of the manufactured component. For example, the heat input can be controlled during manufacturing to avoid harmful metallographic phases by introducing dwell time. In this way, the deposited material properties can reach the requirements of expected mechanic and metallographic specifications. The filer material used for the manufacturing of the blade is a duplex stainless steel (G 22 9 3 NL). Duplex stainless steels are dual-phase material. It confers good mechanical properties and corrosion resistance which is suitable for marine applications. Several test blocks have been built to characterize the AM parts properties. Mechanical properties obtained were superior to the requirements for this application. The micrographic examinations showed a fine austeno-ferritic microstructure and no critical defect was observed in the microstructure. Therefore, the material characterization has validated the use of duplex stainless steel with the WAAM process for the manufacturing of a hollow propeller blade. More details are exposed in the article [11].

## Interest multiple axial redundancies on the equipment

Six degrees of freedom **(**dof) robots are widely used in the industry. There kinematic architecture with 6 nested axes provides a useful redundancy to program the robot movements. For example, the tool orientation can be controlled with more flexibility.

As part of the RAMSSES project, a first 1 m high hollow blade demonstrator has been successfully built with WAAM using a 6 dof robot as exposed in the article [11]. However, this motion system showed some limitations. Firstly, the excessive slope of the trailing edge of the blade induced to use multiple setups in order to control the melting pool orientation. Secondly, the size of the cavity was limited because of the torch accessibility, giving a limited mass reduction (34%) compared to design expectations (around 50%). Finally, this manufacturing strategy is not applicable for monobloc propellers as the part cannot be rotated for the other blades.

Thus, a second hollow propeller blade demonstrator was built on a 8 dof robotic cell, including a 6-axes robot synchronized with a 2-axes positioner. This second demonstrator is bigger than the first one with a 1.5m height. The additional geometric redundancy consequently increased the number of kinematic joint configurations for a given oriented position. However, this freedom gain requires constraining the robot to prevent unexpected configurations or collisions. Furthermore, the kinematic architecture of the robot can introduce singular positions that is harmful to the robot movements. Tool path simulations on a numerical environment is essential to anticipate the robot’s behavior at this stage.

The main interest of an additional positioner is to straighten the tool axes in the case of strong slopes and keep the manufacturing torch vertical. Thus, it simulates flat welding even on sections with high slopes which would not have been possible without an orientation of the torch in the first place. For example, to avoid melt pool collapsing at the trailing edge of the blade, the orientation of the torch is modified to compensate the high slope of this zone. To achieve this, the orientation is gradually changed when the torch reaches the trailing edge by using the positioner, keeping the torch orientation constant with a control of the melting pool orientation relatively to gravity (Fig. 3(a)). Fig. 3(b) shows a view of the hollow blade during the manufacturing process.

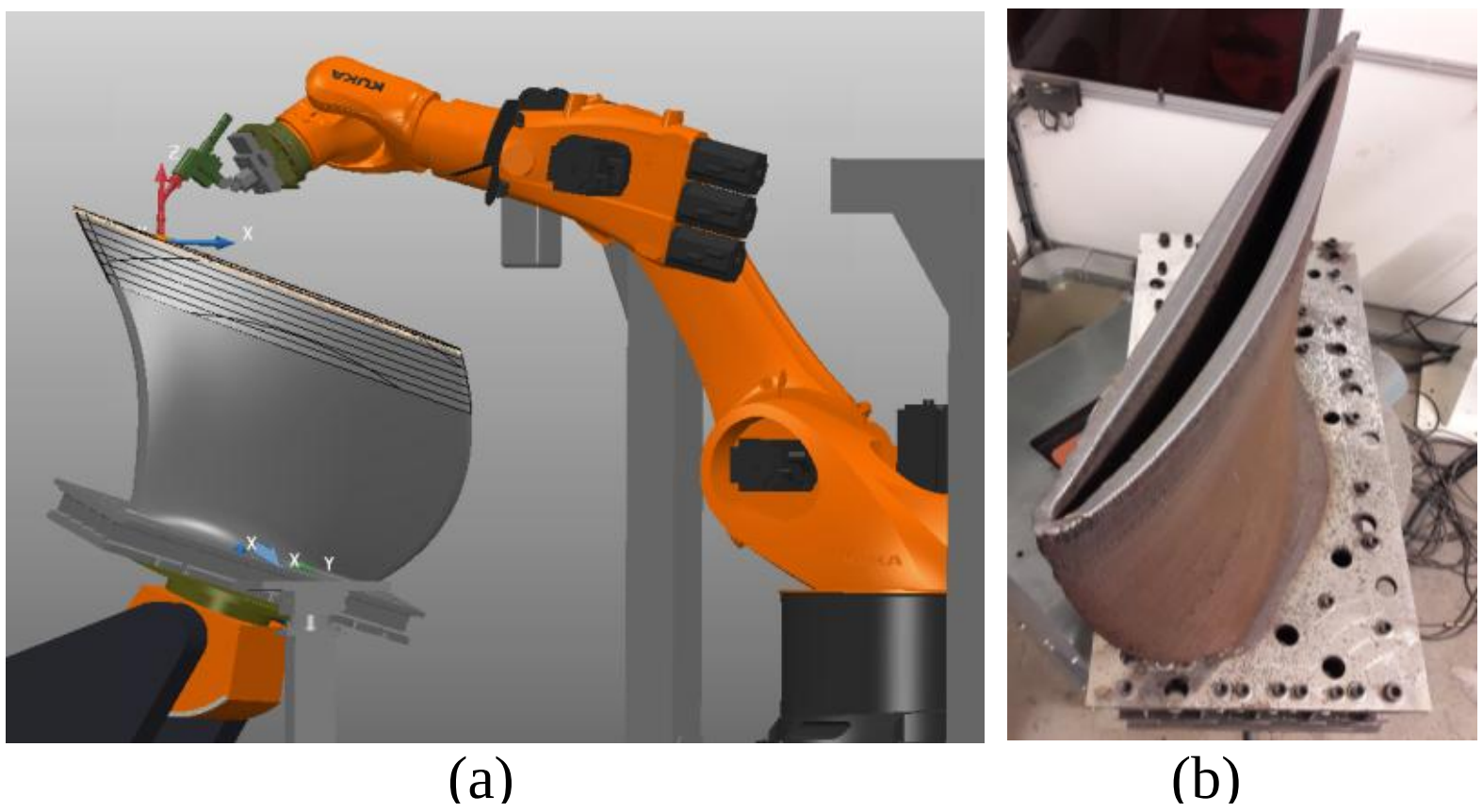

(a) (b)

Fig. 3 (a) Control of the melting pool orientation by using the positioner, (b) View of the hollow blade during manufacturing

This manufacturing strategy employs multiple axial redundancies to build the hollow blade in one setup. Moreover, it is closer to a manufacturing process for a complete propeller as the positioner permits the rotation of the part, better suited for the realization of the other blades.

### Result analysis: Evaluation of the propeller manufactured

Fig. 4 shows the final hollow blade as-fabricated state. By using the manufacturing with 8 dof robotic cell, the closing of the cavity has been improved and the design expectation was met by significant 49% reduction of mass compared to a full blade.

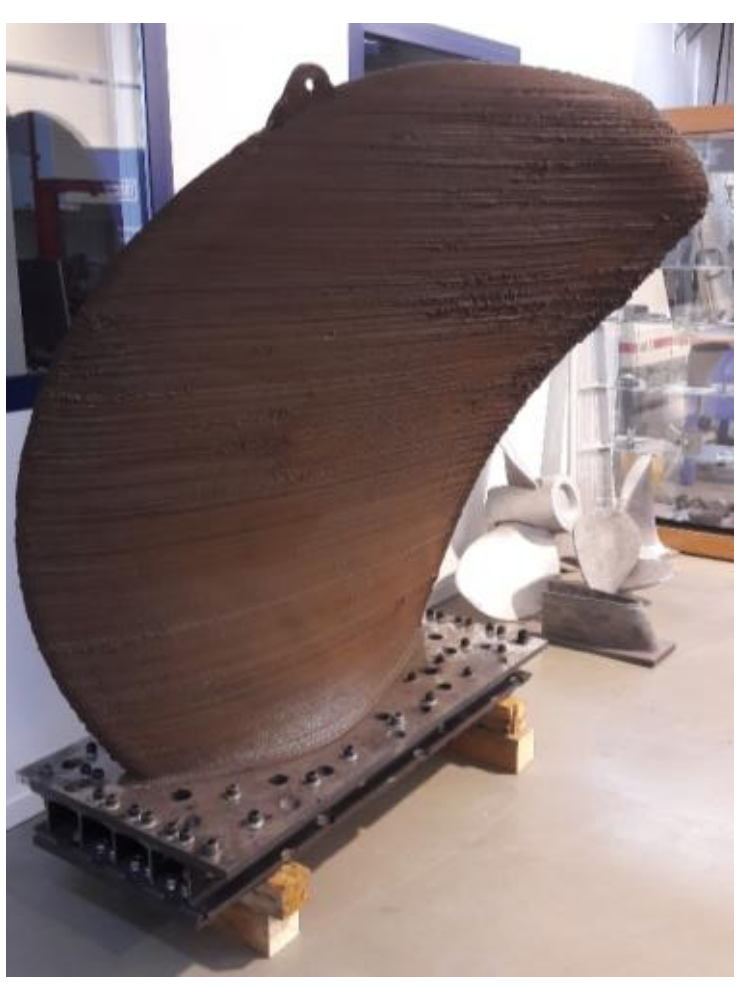

Fig. 4 View of the propeller blade completed

A handheld laser scanner for 3D inspection was used to scan the complete demo blade and its support. Then, a comparison with the CAD model was realized and the results are shown in Fig. 5. These are typical to the expected results for WAAM multi-axis large parts. The maximal deviation is below +/- 1%, which is very good. The little offset of the shape is due to the positioner accuracy and calibration range. The effects of the thermal behavior due to the manufacturing on the accuracy of the WAAM part can also be identified. As a result, these results are very promising as WAAM aims to stand as a near net shape process. The final functional shape will be obtained with finishing operation by high speed machining.

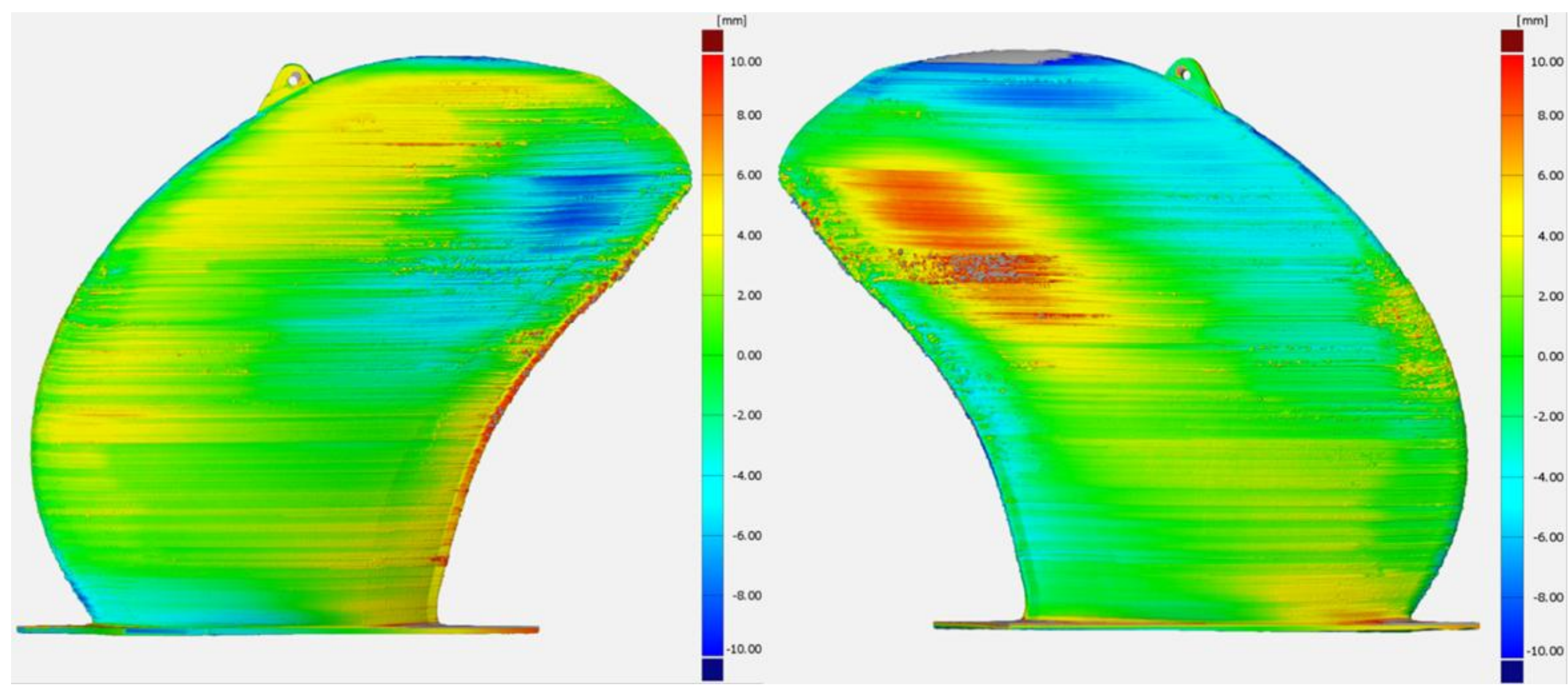

Fig. 5 - Comparison between CAD model and scan of the blade for the both side

## Conclusions

This paper showed the interests of using AM to extend the performances of high value components, by offering new design solutions adapted to this process. The use case studied was a marine propeller blade. The selection of WAAM as manufacturing process enable to develop a hollow blade concept. With AM, the deposition process layer-by-layer provides a better control of the deposited material properties as well as adapting the parameters toolpath during manufacturing. The use of a 8 dof robotic cell, with multiple redundancies, entails complex toolpath programming. However, it effectively improves the manufacturing process as demonstrated by the realization of a hollow propeller blade, which contains complicated geometries to manufacture but necessary for having a good hydrodynamic profile. The development of a methodology by using multi-axial toolpaths allow to manufacture complex areas like high overhanging zones. The application of this methodology helped to realize the hollow propeller blade in one set-up and an expected mass reduction, which would have been impossible without multi-axial toolpath.

## Acknowledgements

The project RAMSSES has received funding under the European Union's Horizon 2020 research and innovation programme under the grant agreement No 723246.